\documentclass[%
 reprint,
superscriptaddress,
 amsmath,amssymb,
 aps,
pra, 
longbibliography,
floatfix ]{revtex4-1}

\usepackage{natbib} 
\usepackage{amsmath} 
\usepackage{graphicx} 
\usepackage[usenames,dvipsnames]{color}  
\usepackage{longtable}
\usepackage[caption=false]{subfig}
\usepackage{multirow}
\usepackage{amsmath}
\usepackage{epstopdf}
\usepackage[flushleft]{threeparttable}
\usepackage[english]{babel}
\usepackage{makecell}
\usepackage[section]{placeins}

\makeatletter
\def\bbl@set@language#1{%
  \edef\languagename{%
    \ifnum\escapechar=\expandafter`\string#1\@empty
    \else\string#1\@empty\fi}%
  \@ifundefined{babel@language@alias@\languagename}{}{%
    \edef\languagename{\@nameuse{babel@language@alias@\languagename}}%
  }%
  \select@language{\languagename}%
  \expandafter\ifx\csname date\languagename\endcsname\relax\else
    \if@filesw
      \protected@write\@auxout{}{\string\select@language{\languagename}}%
      \bbl@for\bbl@tempa\BabelContentsFiles{%
        \addtocontents{\bbl@tempa}{\xstring\select@language{\languagename}}}%
      \bbl@usehooks{write}{}%
    \fi
  \fi}
\newcommand{\DeclareLanguageAlias}[2]{%
  \global\@namedef{babel@language@alias@#1}{#2}%
} \makeatother

\DeclareLanguageAlias{en}{english}

\begin{document}

\title{Using Machine Learning to Identify the Most At-Risk Students in Physics Classes}
\author{Jie Yang}%
\author{Seth DeVore}%
\author{Dona Hewagallage}%
\author{Paul Miller}%
\affiliation{%
Department of Physics and Astronomy, West Virginia University,
Morgantown WV, 26506
}%
\author{Qing X. Ryan}%
    \affiliation{%
        Department of Physics and Astronomy, California State Polytechnic University, Pomona
        CA, 91768
    }%
\author{John Stewart}
\email{jcstewart1@mail.wvu.edu}
\affiliation{%
Department of Physics and Astronomy, West Virginia University,
Morgantown WV, 26506
}%

\date{\today}

\begin{abstract}
Machine learning algorithms have recently been used to predict
students' performance in an introductory physics class. The
prediction model classified students as those likely to receive an
A or B or students likely to receive a grade of C, D, F or
withdraw from the class. Early prediction could better allow the
direction of educational interventions and the allocation of
educational resources. However, the performance metrics used in
that study become unreliable when used to classify whether a
student would receive an A, B or C (the ABC outcome) or if they
would receive a D, F or withdraw (W) from the class (the DFW
outcome) because the outcome is  substantially unbalanced with
between 10\% to 20\% of the students receiving a D, F, or W. This
work presents techniques to adjust the prediction models and
alternate model performance metrics more appropriate for
unbalanced outcome variables. These techniques were applied to
three samples drawn from introductory mechanics classes at two
institutions ($N=7184$, $1683$, and $926$). Applying the same
methods as the earlier study produced a classifier that was very
inaccurate, classifying only 16\% of the DFW cases correctly;
tuning the model increased the DFW classification accuracy to
43\%. Using a combination of institutional and in-class data
improved DFW accuracy to 53\% by the second week of class. As in
the prior study, demographic variables such as gender,
underrepresented minority status, first-generation college student
status, and low socioeconomic status were not important variables
in the final prediction models.

\end{abstract}

\maketitle

\section{Introduction}

Physics courses, along with other core science and mathematics
courses, form key hurdles for Science, Technology, Engineering,
and Mathematics (STEM) students early in their college career.
Student success in these classes is important to improving STEM
retention; the success of students traditionally underrepresented
in STEM disciplines in the core classes may be a limiting factor
in increasing inclusion in STEM fields. Physics Education Research
(PER) has developed a wide range of research-based instructional
materials and practices to help students learn physics
\cite{meltzer2012resource}. Research-based instructional
strategies have been demonstrated to increase student success and
retention \cite{freeman2014active}. While some of these strategies
are easily implemented for large classes, others have substantial
implementation costs. Further, no class could implement all
possible research-based strategies, and some may be more
appropriate for some subsets of students than for others. One
method to better distribute resources to the students who would
benefit the most is to identify at-risk students early in physics
classes. The effective identification of students at risk in
physics classes and the efficacious uses of this classification
represents a promising new research strand in PER.

The need for STEM graduates continues to increase at a rate that
is outstripping STEM graduation rates across American
institutions. A 2012 report from the President's Council of
Advisors on Science and Technology \cite{olson2012engage}
identified the need to increase graduation of STEM majors to avoid
a projected shortfall of one million STEM job candidates over the
next decade. Improving STEM retention has long been an important area of
investigation for science education researchers \cite{rask2010,
chen2013, shaw2010, maltese2011,zhang2004,french2005,
marra2012,hall2015}. Targeting interventions to students
at risk in core introductory science and mathematics courses taken
early in college offers one potential mechanism to improve STEM
graduation rates. In recent years, educational
data mining has become a prominent method of analyzing student
data to inform course redesign and to predict student performance and persistence \cite{baepler2010, baker2009,
papamitsiou2014,dutt2017,romero2010}.

The current study investigates the application of machine learning
algorithms to identify at-risk students. Machine learning and data
science as a whole are growing explosively in many segments of the
economy as these new methods are used to make sense and exploit
the exponentially growing data collected in an increasing online
world. These methods are also being adapted to understand and
improve educational data systems. It seems likely that this
process will accelerate in the near future as universities, in a
challenging financial climate, attempt to retain as many students
as possible. We argue that PER should help shape both the
construction of retention models of physics students and explore
their most effective and most ethical use. The following
summarizes the prior study applying Education Data Mining (EDM)
techniques in physics classes, provides an overview of EDM, and
more specifically an overview of the use of EDM for grade
prediction.

\subsection{Prior Study: Study 1}
This study extends the results of Zabriskie {\it et al.}
\cite{zabriskie2019gpp} which will be referred to as Study 1 in
this work. Study 1 used institutional data such as ACT scores and
college GPA (CGPA) as well as data collected within a physics
class such as homework grades and test scores to predict whether a
student would receive an A or B in the first and second semester
of a calculus-based physics class at a large university. The study
used both logistic regression and random forests to classify
students. Random forest classification using only institutional
variables was 73\% accurate for the first semester class. This
accuracy increased to 80\% by the fifth week of the class when
in-class variables were included. The logistic regression and
random forest classification algorithms generated very similar
results. Study 1 chose to predict A and B outcomes, rather than
the more important A, B, and C outcomes, partially because the
sample was significantly unbalanced. Sample imbalance makes
classification accuracy more difficult to interpret. Study 1
investigated the effect of a number of demographic variables
(gender, underrepresented minority status, and first-generation
status) on grade prediction and found they were not important to
grade classification. These groups (women, underrepresented
minority students, and first-generation students) were very
underrepresented in the sample studied; it was unclear to what
extent the low importance of the demographic variables was caused
by the demographic imbalance of the sample.

\subsection{Research Questions}
This study seeks to extend the application of machine learning algorithms to predict
whether a student will earn a D or F or withdraw (W) from a physics class. In particular, we explore the following research questions:
\begin{enumerate}
    \item[RQ1:] How can machine learning algorithms be applied to predict an unbalanced outcome in a physics class?
    \item[RQ2:] Does classification accuracy differ for underrepresented groups in physics? If so, how and why does it differ?
    \item[RQ3:] How can the results of a machine learning analysis be applied to better understand
    and improve physics instruction?
\end{enumerate}

\subsection{Educational Data Mining}
Educational Data Mining (EDM) can be described as the use of
statistical, machine learning, and traditional data mining methods
to draw conclusions from large educational datasets while
incorporating predictive modeling and psychometric modeling
\cite{romero2010}. In a 2014 meta-analysis of 240 EDM articles by
Pe{\~n}a-Ayala, 88\% of the studies were found to use a statistical and/or
machine learning approach to draw conclusions from the data
presented. Of these studies, 22\% analyzed student behavior, 21\%
examined student performance, and 20\% examined assessments
\cite{pena2014educational}. Pe{\~n}a-Ayala also found that
classification was the most common method used in EDM applied in
42\% of all analyses, with clustering used in 27\%, and regression
used in 15\% of studies.

Educational Data Mining encompasses a large number of statistical
and machine learning techniques with logistic regression, decision
trees, random forests, neural networks, naive Bayes, support
vector machines, and K-nearest neighbor algorithms commonly
applied \cite{romero2008data}. Pe{\~n}a-Ayala's
\cite{pena2014educational} analysis found 20\% of studies employed
Bayes theorem and 18\% decision trees. Decision trees and random
forests are one of the more commonly used techniques in EDM. We use these techniques to investigate our research questions and
explore ways to assess the success of machine learning
algorithms. More information on the fundamentals of these and
other machine learning techniques are readily available through a
number of machine learning texts \cite{james2017,muller2016introduction}.

\subsection{Grade Prediction and Persistence}

While EDM is used for a wide array of purposes, it has often been
used to examine student performance and persistence. One survey by
Shahiri {\it et al.} summarized 30 studies in which student
performance was examined using EDM techniques
\cite{shahiri2015review}. Neural networks and decision trees were
the two most common techniques used in studies examining student
performance with naive Bayes, K-nearest neighbors, and support
vector machines used in some studies. A study by Huang and Fang
examined student performance on the final exam for a
large-enrollment engineering course using measurements of college
GPA, performance in 3 prerequisite math classes as well as Physics
1, and student performance on in-semester examinations
\cite{huang2013predicting}. They analyzed the data using a large
number of techniques commonly used in EDM and found relatively
little difference in the accuracy of the resulting models. Study 1
also found little difference in the performance of machine
learning algorithms in predicting physics grades. Another study
examining an introductory engineering course by Marbouti {\it et
al.} used an array of EDM techniques to predict student grade
outcomes of C or better \cite{marbouti2016models}. They used
in-class measures of student performance including homework, quiz,
and exam 1 scores and found that logistic regression provided the
highest accuracy at 94\%. A study by Macfadyen  and Dawson
attempted to identify students at risk of failure in an
introductory biology course \cite{macfadyen2010mining}. Using
logistic regression they were able to identify students failing
(defined as having a grade of less than $50\%$) with 81\%
accuracy. With the goal of improving STEM retention, many universities are taking a rising interest in using EDM techniques for grade and persistence prediction in STEM classes
\cite{bin2013overview}.

The use of machine learning techniques in physics classes has only
begun recently. In addition to Study 1, random forests were used in a 2018
study by Aiken {\it et al.} to predict student persistence as
physics majors and identify the factors that are predictive of
students either remaining physics majors or becoming engineering
majors \cite{aiken2018modeling}.

\section{Methods}
\label{sec:methods}

\def\tm{$\times$}

\renewcommand{\tabcolsep}{2mm}
\begin{table*}[!htb]
    \caption{\label{tab:variables} Full list of variables.}
    \begin{center}

        \begin{tabular}{lccccl}\hline

            Variable  &\multicolumn{3}{c}{Sample}&  Type & Description\\
            &1&2&3&&\\\hline
            \multicolumn{6}{c}{Institutional Variables}\\\hline
            Gender    &\tm&\tm&\tm&Dichotomous&  Does the student identify as a man or a women?                                                      \\
            URM       &\tm&\tm&\tm&Dichotomous&  Does the student identify as an underrepresented minority? \\
            FirstGen  &\tm&\tm&\tm&Dichotomous&  Is the student a first-generation college student?                                 \\
            CalReady &\tm&\tm&&Dichotomous&   Is the student ready for calculus?  \\
            SES      &&&\tm&Dichotomous&  Does the student qualify for a Pell grant?                                 \\
            CmpPct    &\tm&\tm&&Continuous&  Percentage of credit hours attempted that were completed. \\
            CGPA      &\tm&\tm&\tm&Continuous&  College GPA at the start of the course.                                                       \\
            STEMCls   &\tm&\tm&&Continuous&  Number of STEM classes completed at the start of the course.       \\
            HrsCmp    &\tm&\tm&&Continuous&  Total credits hours earned at the start of the course.                                                                \\
            HrsEnroll &\tm&\tm&&Continuous&  Current credits hours enrolled at the start of the course.                                                           \\
            HSGPA     &\tm&\tm&\tm&Continuous&  High school GPA.                                                                              \\
            ACTM      &\tm&\tm&\tm&Continuous&  ACT/SAT mathematics percentile score.                                                                      \\
            ACTV      &\tm&\tm&&Continuous&  ACT/SAT verbal percentile score.                                                                    \\
            APCredit  &\tm&\tm&&Continuous&  Number of credits hours received from AP courses.                                                     \\
            TransCrd  &\tm&\tm&&Continuous&  Number of credits hours received from transfer courses.                                             \\\hline
            \multicolumn{6}{c}{In-Class Variables}\\\hline
            Clicker  &&\tm&&Continuous&  Average clicker score graded for participation.                                             \\
            Homework  &&\tm&&Continuous&  Homework average.                                             \\
            TestAve  &&\tm&&Continuous&  Average for the first or the first and second exam.                                             \\
            Pretest Participation  &&\tm&&Dichotomous&  Was the pretest taken?                                             \\
            Pretest Score  &&\tm&&Continuous&  FMCE pretest score.                                             \\\hline
        \end{tabular}
    \end{center}
\end{table*}
\renewcommand{\tabcolsep}{2pt}

\subsection{Sample}
\label{sec:context}

This study used three samples drawn from the introductory calculus-based
physics classes at two institutions.

Sample 1 and 2 were collected in the
introductory, calculus-based mechanics course (Physics 1) taken by
physical science and engineering students at a large eastern
land-grant university (Institution 1) serving approximately 21,000 undergraduate students. The
general university undergraduate population had ACT scores ranging
from 21 to 26 (25th to 75th percentile) \cite{usnews}. The overall
undergraduate demographics were 80\% White, 4\% Hispanic, 6\%
international, 4\% African American, 4\% students reporting
two or more races, 2\% Asian, and other groups each with 1\% or less \cite{usnews}.

Sample 1 was drawn from institutional records and includes all
students who completed Physics 1 from 2000 to 2018, for a sample
size of 7184. Over the period studied, the instructional
environment of the course varied widely, and as such, the result
for this sample may be robust to pedagogical variations.  Prior to
the spring 2011 semester, the course was presented traditionally
with multiple instructors teaching largely traditional lectures
and students performing cookbook laboratory exercises. In spring
2011, the department implemented a Learning Assistant (LA) program
\cite{otero2010physics} using the
 {\it Tutorials in Introductory Physics}
\cite{mcdermott1998}.  In fall 2015, the program was modified
because of a funding change with LAs assigned to only a subset of
laboratory sections. The Tutorials were replaced with open source
materials \cite{opensource} which lowered textbook costs to
students and allowed full integration of the research-based
materials with laboratory activities.

Sample 2 was collected from the fall 2016 to the spring 2019 semester when the instructional environment
was stable, for a sample size of 1683. The same institutional data were collected and the sample also included
a limited number of in-class performance measures: clicker average, homework average,
Force and Motion Conceptual Evaluation (FMCE) pretest score, FMCE pretest participation, and the score on in-semester examinations. A more detailed explanation of these variables will be provided in the next section.

Sample 3 was collected at a primarily undergraduate and
Hispanic-serving university (Institution 2) with approximately
26,000 students in the western US. Fifty percent of the general
undergraduate population had ACT scores in the range 19 to 27. The
demographics of the general undergraduate population were 46\%
Hispanic, 21\% Asian, 16\% White, 6\% International, 4\% two or
more races, 3\% African American, 3\% unknown, with other races
1\% or less \cite{usnews}. The sample was collected in the
introductory calculus-based mechanics class for all four quarters
of the 2017 calendar year. This class also primarily serves
physical science and engineering students. The course was taught
in multiple sections each quarter with multiple different
instructors. The pedagogical style varied greatly with some
instructors giving traditional lectures and some teaching using
active-learning methods.

\subsection{Variables}

The variables used in this study were drawn from institutional
records and from data collected within the classes and are shown in Table \ref{tab:variables}. Two types of
variables were used: two-level dichotomous variables and continuous
variables. A few variables require additional explanation. The
variable CalReady measures the student's math-readiness. Calculus
1 is a pre-requisite for Physics 1. For the vast majority of
students in Physics 1, the student's four-year degree plans assume
the student enrolls in Calculus 1 their first semester at the
university. These students are considered ``math ready.'' A
substantial percentage of the students at Institution 1
are not math ready.  The variable STEMCls captures the number of
STEM classes completed before the start of the
course studied. STEM classes include mathematics, biology,
chemistry, engineering, and physics classes.

For all samples, demographic information was also collected from institutional
records. Students
were considered first generation if neither of their parents
completed a four-year degree. A student was classified as an underrepresented
minority student (URM) if they identified as
Hispanic or reported a race other than White or Asian. Gender was
also collected from university records; for the period studied
gender was recorded as a binary variable. While
not optimal, this reporting is consistent with the use of gender
in most studies in PER; for a more nuanced discussion of gender
and physics see Traxler {\it et al.}
\cite{traxler_enriching_2016}.

For Sample 2, in-class data were also available on a weekly basis. This data included
clicker scores (given for participation points), homework averages, test scores, and
a conceptual pretest score (PreScore) using the FMCE \cite{thornton1998}.
Students not in attendance on the day the FMCE was given received a zero; whether students completed
the FMCE was captured by the dichotomous variable (PreTaken) which is one if the test was taken, zero otherwise.

For Sample 3, socioeconomic status (SES) was measured by whether the students qualified for a federal Pell grant. A student is
eligible for a Pell grant if their family income is less than $\$50,000$ US dollars; however, most Pell grants are awarded to
students with family incomes less than $\$20,000$ \cite{pell}.

\subsection{Random Forest Classification Models}

This work employs the random forests machine learning algorithm to
predict students' final grade outcomes in introductory physics.
Random forests are one of many machine learning classification
algorithms. Study 1 reported that most machine learning algorithms
had similar performance when predicting physics grades. A
classification algorithm seeks to divide a dataset into multiple
classes. This study will classify students as those who will
will receive an A, B, or C (ABC students) and students who
will receive a D or F or withdraw (W) (DFW students).

To understand the performance of a classification algorithm, the
dataset is first divided into test and training datasets. The
training dataset is used to develop the classification model, to
train the classifier. The test dataset is then used to
characterize the model performance. The classification model is
used to predict the outcome of each student in the test dataset;
this prediction is compared to the actual outcome. Section
\ref{sec:pred} discusses performance metrics used to characterize
the success of the classification algorithm. For this work, 50\%
of the data were included in the test dataset and 50\% in the
training dataset. This split was selected to maintain a
substantial number of underrepresented students in both the test
and training datasets.

The random forest algorithm uses decision trees, another machine
learning classification algorithm. Decision trees work by
splitting the dataset into two or more subgroups based on one of
the model variables. The variable selected for each split is
chosen to divide the dataset into the two most homogeneous subsets
of outcomes possible, that is, subsets with a high percentage of
one of the two classification outcomes. The variable and the
threshold for the variable represents the decision for each node
in the tree. For example, one node may split the dataset using the
criteria (the decision) that a student's college GPA is less than
3.2. The process continues by splitting the subsets forming the
decision tree until each node contains only one of the two
possible outcomes. Decision trees are less susceptible to
multicollinearity than many statistical methods common in PER such
as linear regression \cite{breiman1984classification}.

Random forests  extend the decision tree algorithm by growing many
trees instead of a single tree. The ``forest'' of decision trees
is used to classify each instance in the data; each tree ``votes''
on the most probable outcome. The decision threshold determines
what fraction of the trees must vote for the outcome for the
outcome to be selected as the overall prediction of the random
forest. Random forests use bootstrapping to prevent one variable
from being obscured by another variable. Bootstrapping is a
statistical method where multiple random subsets of a dataset are
created by sampling with replacement. Individual trees are grown
on $Z$ subsamples generated by sampling the training data set with
replacement using a subset of size $m=\sqrt{k}$ of the variables,
where $k$ is the number of independent variables in the model
\cite{hastie2009}. This method ensures the trees are not
correlated and that the stronger variables do not overwhelm weaker
variables \cite{james2017}. The ``randomForest'' package in ``R''
was used for the analysis. The Supplemental Material contains an
example of random forest code in R \cite{supp}.

\subsection{Performance Metrics} \label{sec:pred}

The confusion matrix \cite{fawcett2006introduction} as shown in
Table \ref{confmat} summarizes the results of a classification algorithm
and is the basis for calculating most model performance metrics. To construct
the confusion matrix, the classification model developed from the training
dataset is used to classify students in the test dataset. The confusion matrix
categorizes the outcome of this classification.

\begin{table}[!htb]
    \caption{\label{confmat} Confusion Matrix}
    \begin{tabular}{c | c | c }
        &Actual Negative& Actual Positive \\\hline
        Predicted Negative& True Negative (TN) & False Negative (FN) \\\hline
        Predicted Positive& False Positive (FP)& True Positive
        (TP)\\
    \end{tabular}

\end{table}

For classification, one of the dichotomous outcomes is selected as
the positive result. In the current study, we use the DFW outcome
as the positive result. This choice was made because some of the
model performance metrics focus on the positive results and we
feel that most instructors would be more interested in accurately
identifying students at risk of failure.

From the confusion matrix, many performance metrics can be
calculated. Study 1 reported the overall classification accuracy, the
fraction of correct predictions, shown in Eqn. \ref{eqn:acc}
\begin{equation}
\label{eqn:acc} \mbox{Overall Accuracy} = \frac{\mbox{TN} +
    \mbox{TP}}{N_{\mathrm{ test}}}
\end{equation}
where $N_{\mathrm{ test}}=\mbox{TP+TN+FP+FN}$ is the size of the
test dataset.

The true positive rate (TPR) and the true negative rate (TNR)
characterize the rate of making accurate predictions of either the
DFW or the ABC outcome. The DFW accuracy is the fraction of the
actual DFW cases that are classified as DFW (Eqn \ref{eqn:sen}) in the test dataset.
\begin{equation}
\label{eqn:sen} \mbox{DFW Accuracy} = \mbox{TPR}
    = \frac{\mbox{TP}}{
        \mbox{TP}+
        \mbox{FN}
        }
\end{equation}
ABC accuracy is the fraction of the actual ABC cases that are
classified as ABC (Eqn \ref{eqn:spec}).
\begin{equation}
\label{eqn:spec} \mbox{ABC Accuracy} =\mbox{TNR}=
\frac{\mbox{TN}}{
    \mbox{TN}+
    \mbox{FP}
    }
\end{equation}
DFW accuracy is called ``sensitivity'' or ``recall'' in machine learning; ABC accuracy is ``specificity.''

ABC and DFW accuracy can be adjusted by changing the strictness of
the classification criteria. If the model classifies even the only
slightly promising cases as DFW, it will probably classify most
actual DFW cases as DFW producing a high DFW accuracy. It will also
make a lot of mistakes; the DFW precision or the positive
predictive value (PPV) captures the rate of making correct
predictions and is defined as the fraction of the DFW predictions which are correct (Eqn. \ref{eqn:prec}).
\begin{equation}
\label{eqn:prec} \mbox{DFW Precision} =\mbox{PPV}=
\frac{\mbox{TP}}{
    \mbox{TP}+
    \mbox{FP}
    }
\end{equation}
DFW precision is called ``precision'' or ``positive predictive value'' in machine learning.

This study seeks models that balance DFW accuracy and
precision; however, the correct balance for a given application
must be selected based on the individual features of the
situation. If there is little cost and no risk to an intervention,
then optimizing for higher DFW accuracy might be the correct
choice to identify as many DFW students  as possible. If the
intervention is expensive or carries risk, optimizing the DFW
precision so that most students who are given the intervention are
actually at risk might be more appropriate.

Beyond simply evaluating the overall performance of a
classification algorithm, we would like to establish how much
better the algorithm performs than pure guessing. For example, Sample 1 is substantially unbalanced between the DFW and ABC
outcomes with 88\% of the students receiving an A, B, or C. If a
classification method guessed that all student would receive an A,
B, or C, then the classifier would have an overall accuracy of
$0.88$; therefore, overall accuracy would not be a useful metric to characterize model performance in this case.

In order to provide a more complete picture of model performance, additional performance metrics were explored. Cohen's kappa,
$\kappa$, measures agreement among observers \cite{cohen}
correcting for the effect of pure guessing as shown in Eqn.
\ref{eqn:kappa}.
\begin{equation}
\label{eqn:kappa} \kappa = \frac{p_0-p_e}{1-p_e}
\end{equation}
where $p_0$ is the observed agreement and $p_e$ is agreement by chance. Fit criteria
have been developed for $\kappa$ with
$\kappa$ less than $0.2$ as poor agreement, 0.2 to 0.4 fair
agreement, 0.4 to 0.6 moderate agreement, 0.6 to 0.8  good
agreement, and 0.8 to 1.0 excellent agreement between observers
\cite{altman1990practical}.

The Receive Operating Characteristic (ROC) curve (originally
developed to evaluate radar) plots the true positive rate (TPR)
against the false positive rate (FPR). The Area Under the Curve
(AUC) is a measure of the model's discrimination between the two
outcomes; AUC is the integrated area under the ROC curve. For a
classifier that uses pure guessing, the ROC curve is a straight
line between (0,0) and (1,1) and the AUC is $0.5$. An AUC of 1.0
represents perfect discrimination
\cite{hosmer2013applied,fawcett2006introduction}. Hosmer
\textit{et al.} \cite{hosmer2013applied}  suggest an AUC threshold
of $0.80$ for excellent discrimination.

\renewcommand{\tabcolsep}{2mm}
\begin{table*}[!htb]
        \centering
        \begin{tabular}{| l| c c ccc| }\hline
            &N&Physics Grade&ACT Math \%&HSGPA & CGPA\\\hline
            Overall     &7184       &$2.70\pm1.3$   &$79\pm14$      &$3.71\pm0.5$       &$3.18\pm0.5$\\\hline
            ABC Students&6337       &$3.05\pm0.8$   &$80\pm14$      &$3.75\pm0.4$       &$3.25\pm0.5$\\
            DFW Students &847        &$0.05\pm0.9$   &$73\pm15$      &$3.43\pm0.5$       &$2.65\pm0.5$\\\hline
            Women       &1270       &$2.83\pm1.2$   &$79 \pm 14$    & $ 3.94\pm 0.4$    & $3.38 \pm 0.5$\\
            Men         &5914       &$2.67 \pm 1.3$ &$79 \pm 14$    &$3.66 \pm 0.5$     &$ 3.14\pm 0.5$\\\hline
            URM         &388        &$2.42 \pm 1.3$ &$73 \pm 17$  &$3.53 \pm 0.5$     &$ 3.03\pm 0.6$\\
            Not URM     &6796       &$2.71 \pm 1.3$ &$80 \pm 14$    &$3.72 \pm 0.5$     &$ 3.19\pm 0.5$\\\hline
            First Gen.  &815        &$2.66\pm1.3$   &$77\pm15$      &$3.72\pm0.5$       &$3.15\pm0.5$\\
            Not First Gen.&6369     &$2.70\pm1.3$   &$80\pm14$      &$3.71\pm0.5$       &$3.18\pm0.5$\\\hline
        \end{tabular}
    \caption{\label{tab:demo} Descriptive statistics for Sample 1. All values are the mean $\pm$ the standard deviation. }
\end{table*}
\renewcommand{\tabcolsep}{2pt}

\renewcommand{\tabcolsep}{2mm}
\begin{table*}[!htb]
    \centering
    \begin{tabular}{| l| c| c |c|c|c| c|c|c|c|}\hline
        &N&Physics Grade&SAT Math \%& HSGPA & CGPA\\\hline
        Overall     &926       &$2.34\pm1.2$   &$75\pm18$     &$3.66\pm0.4$       &$3.10\pm0.6$\\\hline
        ABC Students&740       &$2.83\pm0.8$   &$77\pm17$     &$3.70\pm0.3$      &$3.20\pm0.5$\\
        DFW Students &186        &$0.39\pm0.5$   &$68\pm19$    &$3.49\pm0.4$       &$2.70\pm0.5$\\\hline
        Women       &259       &$2.21\pm1.2$  &$71 \pm 19$    & $ 3.70\pm 0.3$    & $3.13 \pm 0.5$\\
        Men         &667       &$2.39 \pm 1.2$ &$77 \pm 17$   &$3.64 \pm 0.4$    &$ 3.09\pm 0.6$\\\hline
        URM         &396        &$2.13 \pm 1.3$ &$68 \pm 19$  &$3.64 \pm 0.4$     &$ 3.02\pm 0.6$\\
        Not URM     &530       &$2.49 \pm 1.2$ &$81 \pm 14$   &$3.67 \pm 0.3$     &$ 3.16\pm 0.5$\\\hline
        First Gen.  &440        &$2.18\pm1.2$   &$70\pm19$   &$3.63\pm0.4$       &$3.03\pm0.6$\\
        Not First Gen.&486    &$2.49\pm1.2$   &$80\pm15$     &$3.68\pm0.3$       &$3.16\pm0.6$\\\hline
        Low SES  &351        &$2.26\pm1.2$   &$71\pm19$      &$3.65\pm0.4$       &$3.06\pm0.6$\\
        Not Low SES&575     &$2.39\pm1.2$   &$78\pm16$      &$3.67\pm0.3$      &$3.12\pm0.6$\\\hline
    \end{tabular}
    \caption{\label{tab:demoSample3} Descriptive statistics for Sample 3. All values are the mean $\pm$ the standard deviation. }
\end{table*}
\renewcommand{\tabcolsep}{2pt}

\subsection{Model Tuning and Validation}
We will find that the random forest classification models have poor performance predicting
whether a student will receive a D, F, or W using the default parameters of the model. To
improve performance, the models are tuned by adjusting the decision threshold. The imbalance of
both the outcome variable and some of the demographic variables must also be investigated
to verify that the models are valid and the conclusions are reliable. This process is described in
detail the Supplemental Material \cite{supp}.

\section{Results}

General descriptive statistics are shown in Table \ref{tab:demo}
and \ref{tab:demoSample3} for Samples 1 and 3 respectively. The
descriptive statistics for Sample 2 are similar to Sample 1 and
are presented in the Supplemental Material \cite{supp}. The
dichotomous outcome variable divides each sample into two subsets
with different academic characteristics. The dichotomous
independent variables further divide the subsets defined by the
outcome variables. The overall demographic composition of the
sample is shown for each sample in the Supplemental Material
\cite{supp}.

\subsection{Classification Models}

To explore the classification of DFW students, multiple classification models
were constructed for each sample. To allow comparison, each model was tuned
so that the DFW accuracy and DFW precision was approximately equal. Table \ref{tab:finalfit}
shows the overall model fit for all
samples. Each sample is discussed separately.

\renewcommand{\tabcolsep}{2mm}
\begin{table*}[!htb]
    \caption{\label{tab:finalfit} Model performance parameters. Values represent the mean $\pm$ the standard
    deviation. \label{tab:results}}
    \begin{center}

        \begin{tabular}{|l|cccccc|}\hline
Model&Overall    & DFW       & ABC &DFW               & $\kappa$             & AUC          \\
&Accuracy    & Accuracy       & Accuracy &Precision               &             &           \\\hline
\multicolumn{7}{|c|}{Sample 1 ($N=7184$)}\\\hline
Default&$0.89\pm 0.00$      &  $0.16\pm 0.02$   & $0.98\pm 0.00$ & $0.57\pm 0.04$    & $0.21\pm 0.02$    & $0.57\pm 0.01$ \\\hline
Overall&$0.87\pm 0.01$    &  $0.43\pm 0.02$   & $0.93\pm 0.01$ & $0.44\pm 0.02$    & $0.36\pm 0.02$    & $0.68\pm 0.01$ \\\hline
Female Students&$0.90\pm 0.01$    &  $0.38\pm 0.05$   & $0.96\pm 0.01$ &$0.49\pm 0.06$     & $0.37\pm 0.05$    & $0.67\pm 0.03$               \\\hline
URM Students&$0.80\pm 0.02$    &  $0.48\pm 0.07$   & $0.86\pm 0.02$ &$0.40\pm 0.06$     & $0.32\pm 0.06$    & $0.67\pm 0.04$            \\\hline
First-Generation Students&$0.87\pm 0.01$    &  $0.44\pm 0.06$   & $0.92\pm 0.01$ &$0.42\pm 0.06$     & $0.35\pm 0.05$    & $0.68\pm 0.03$  \\\hline
Restricted&$0.85\pm 0.01$    &  $0.36\pm 0.02$   & $0.91\pm 0.01$ & $0.36\pm 0.02$    & $0.28\pm 0.02$    & $0.64\pm 0.01$ \\\hline
\multicolumn{7}{|c|}{Sample 2 ($N=1683$)}\\\hline
Institutional&$0.90\pm 0.01$    &  $0.50\pm 0.05$   & $0.95\pm 0.01$ & $0.50\pm 0.04$    & $0.45\pm 0.04$    & $0.73\pm 0.02$ \\\hline 
In-Class Only - Week 1&$0.88\pm 0.01$    &  $0.37\pm 0.05$   & $0.94\pm 0.02$ & $0.38\pm 0.05$    & $0.31\pm 0.04$    & $0.65\pm 0.02$ \\\hline 
Institutional and In-Class Week 1&$0.91\pm 0.01$    &  $0.53\pm 0.05$   & $0.95\pm 0.01$ & $0.53\pm 0.04$    & $0.48\pm 0.04$    & $0.74\pm 0.02$ \\\hline 
In-Class Only Week 2& $0.89\pm 0.01$    &  $0.42\pm 0.05$   & $0.94\pm 0.01$ & $0.43\pm 0.05$    & $0.36\pm 0.04$    & $0.68\pm 0.02$ \\\hline 
Institutional and In-Class Week 2&$0.91\pm 0.01$    &  $0.56\pm 0.05$   & $0.95\pm 0.01$ & $0.55\pm 0.04$    & $0.51\pm 0.04$    & $0.76\pm 0.02$ \\\hline 
In-Class Only Week 5& $0.92\pm 0.01$    &  $0.54\pm 0.06$   & $0.95\pm 0.01$ & $0.54\pm 0.05$    & $0.49\pm 0.04$    & $0.74\pm 0.03$ \\\hline 
Institutional and In-Class Week 5&$0.93\pm 0.01$    &  $0.59\pm 0.05$   & $0.96\pm 0.01$ & $0.60\pm 0.05$    & $0.55\pm 0.04$    & $0.78\pm 0.04$ \\\hline 
In-Class Only Week 8 & $0.94\pm 0.01$    &  $0.66\pm 0.05$   & $0.96\pm 0.01$ & $0.65\pm 0.05$    & $0.62\pm 0.04$    & $0.81\pm 0.03$ \\\hline 
Institutional and In-Class Week 8&$0.94\pm 0.01$    &  $0.68\pm 0.05$   & $0.97\pm 0.01$ & $0.68\pm 0.04$    & $0.65\pm 0.04$    & $0.82\pm 0.02$ \\\hline 
\multicolumn{7}{|c|}{Sample 3 ($N=926$)}\\\hline
Overall&$0.74\pm 0.02$    &  $0.37\pm 0.05$   & $0.84\pm 0.03$ & $0.37\pm 0.03$    & $0.21\pm 0.04$    & $0.61\pm 0.02$ \\\hline
Female Students&$0.70\pm 0.02$    &  $0.40\pm 0.08$   & $0.79\pm 0.04$ & $0.38\pm 0.05$    & $0.19\pm 0.06$    & $0.60\pm 0.03$\\\hline
URM Students&$0.67\pm 0.03$    &  $0.41\pm 0.09$   & $0.76\pm 0.05$ &$0.37\pm 0.05$     & $0.16\pm 0.06$    & $0.58\pm 0.04$            \\\hline
First-Generation Students&$0.72\pm 0.02$    &  $0.45\pm 0.07$   & $0.80\pm 0.03$ &$0.43\pm 0.04$     & $0.25\pm 0.06$    & $0.63\pm 0.03$  \\\hline
Low SES Students&$0.72\pm 0.03$    &  $0.35\pm 0.09$   & $0.82\pm 0.05$ & $0.36\pm 0.06$    & $0.17\pm 0.07$    & $0.58\pm 0.04$\\\hline
        \end{tabular}
    \end{center}
\end{table*}

\renewcommand{\tabcolsep}{2pt}

\subsubsection{Sample 1}
\label{sec:amp1}

Sample 1 was first analyzed using the default decision threshold
for the randomForest package in ``R'' where 50\% of the trees must
vote for the outcome to be selected. This was the threshold used
in Study 1. This result is shown as the ``Default'' model in Table
\ref{tab:finalfit}. The model has very poor DFW accuracy with only
16\% of the DFW students identified. It also has fairly poor
$\kappa$ and AUC. This poor performance results from the
unbalanced DFW outcome where only 12\% of the students receive a
D, F, or W. This model was tuned to produce the ``Overall'' model
by adjusting the decision threshold as shown in the Supplemental
Material \cite{supp}. A threshold of 32\% of trees voting for the
DFW classification produced the Overall model which balanced DFW
accuracy and precision. This model substantially improved DFW
accuracy to 43\% at the expense of lower DFW precision and had
substantially better $\kappa$ and AUC; $\kappa=0.36$ represented
fair agreement; however, the AUC value of 0.68 was well below
Hosmer's threshold of 0.80 for excellent discrimination.

The classification model constructed on the full training dataset
was then used to classify each demographic subgroup in the test
dataset to determine if a model trained on a sample composed
predominantly of majority students would be accurate for other
students. The $\kappa$ and AUC of the models classifying women,
URM students, and first-generation students were very similar.
Some, but not extreme, variation was measured for DFW accuracy and
precision. The overall classifier had lower DFW accuracy for women
and higher accuracy for URM students (with corresponding changes
in precision). This may indicate that it would be productive to
tune the models separately for different demographic groups.

Finally, the model labeled ``Restricted'' was constructed using
only a subset of variables similar to those available for Sample
3. Sample 3 contained institutional variables that are commonly
supplied with a demographic data request to institutional records;
Sample 1 also included variables such as STEMCls which may be of
particular interest for prediction of the outcomes of physics
students and variables such as the percentage of classes completed
that may be of particular importance in DFW classification. As one
might expect, the Restricted model using fewer variables performed
more weakly than the Overall model with DFW accuracy reduced by
7\%.

\subsubsection{Sample 2}
Sample 2 contained the same institutional variables as Sample 1,
but also included in-class data such as homework grades and
clicker grades which were available on a weekly basis. While the
institutional data would require a data request to institutional
research at most institutions, the in-class variables should be
available to most physics instructors. Table \ref{tab:finalfit}
shows the progression of DFW accuracy and precision as the class
progresses.

\begin{figure*}[!htb]
    \centering
    \includegraphics[width=6in]{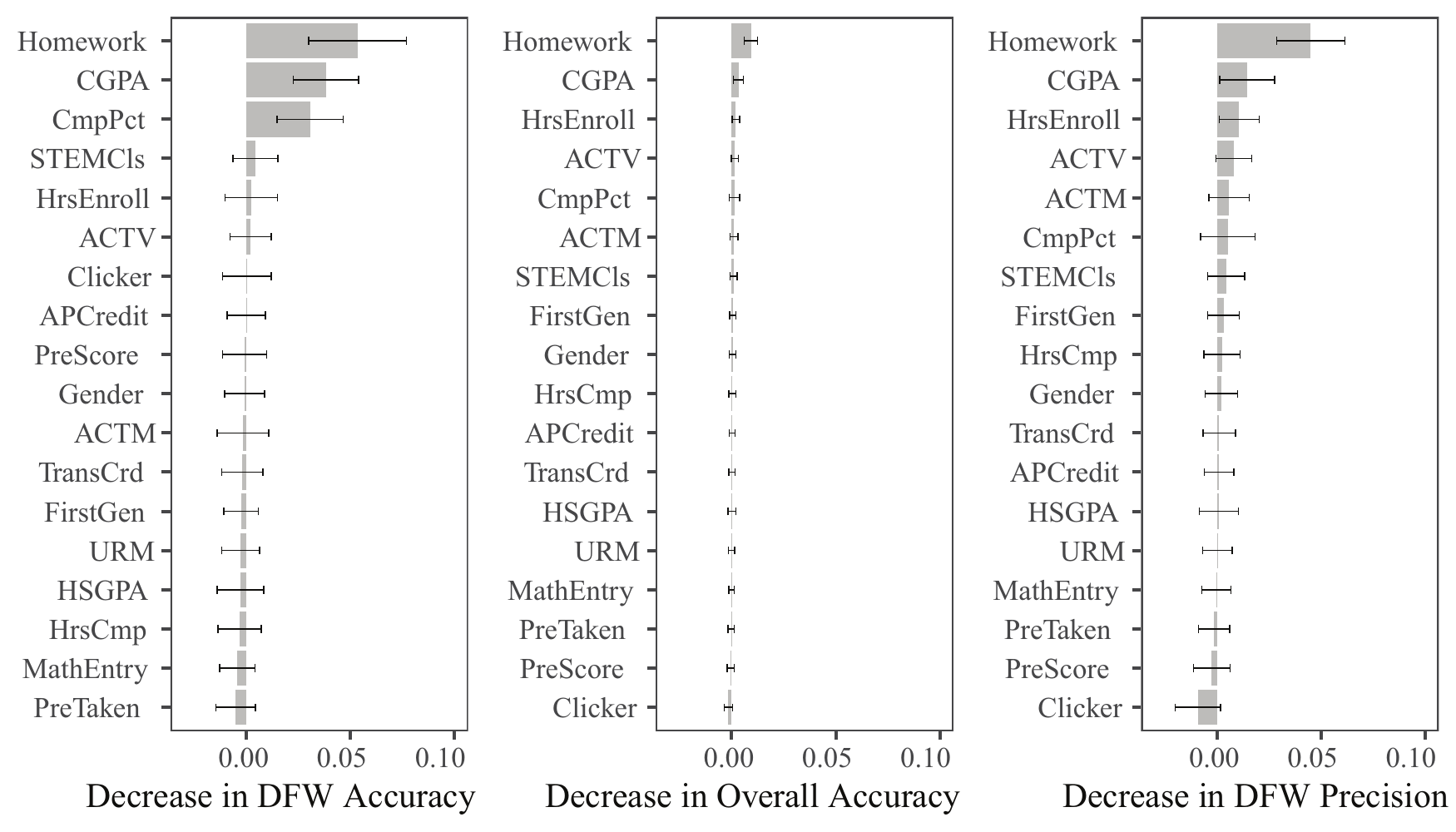}
    \caption{Variable importance of the optimized model predicting DFW for Sample 2 using institutional data and
    data available in-class at the end of week 2. Error bars are
    one standard deviation in length. \label{fig:finalvar}   }
\end{figure*}

A model using only the institutional variables was first
constructed to determine how well DFW students could be identified
using only variables available before the semester begins. This
model (Institutional) had superior performance characteristics to
the Overall model of Sample 1 which used the same variables and a
larger sample collected over a longer time period. The improved
performance quite possibly was the result of Sample 1 averaging
over many instructional environments while Sample 2 contained data
from a single instructional design. This suggests that limiting
the data used for the classifier to the current implementation of
a course may produce superior results, even with lower sample
size.

The performance of models using only the in-class data easily
available to instructors consistently performed more weakly than
those which mixed in-class and institutional data. The
in-class-only models  improved as the class progressed and became
better than the model including only institutional data after the
first test was given in week 5. The in-class-only model was
substantially better than the institutional model after the second
test was given in week 8. As such, if the goal of a classification
algorithm is to predict student outcomes well into the class, only
in-class data is needed.

The models combining in-class and institutional data added
surprisingly little predictive power to the institutional model,
particularly early in the class. This further supports the need to
access a rich set of institutional data for accurate
classification early in a class and suggests predictions made
using only institutional data will not be substantially modified
using in-class data until the first test is given.

\subsubsection{Sample 3}

As shown in Table \ref{tab:variables}, Sample 3 contains many
fewer variables than Sample 1. The classification model for Sample
3 had lower DFW accuracy and precision than similar models for
Samples 1 and 2. Restricting the variable set of Sample 1 to be
approximately that of Sample 3 (the Reduced model) produced a
classifier with similar properties to that of Sample 3. The
difference in classification accuracy, therefore, seems to be the
result of the difference in the variables available and not the
difference in sample size or differences between the universities.

The student population of Sample 3 is substantially more diverse than that of Sample 1 or 2. Model performance predicting only the outcomes of minority demographic subgroups was approximately that of the overall model performance with somewhat lower variation than Sample 1. This suggests that the differences in model performance for demographic subgroups observed in Sample 1 were not a result of the low representation of those groups in the sample. Low SES students were also analyzed separately; the model performance for low SES students was similar to the overall model performance.

\subsection{Variable Importance}
\label{sec:varimp}

Once constructed, classification models can provide physics instructors and
departments a much more nuanced picture of student risk and provide tools to better
serve their students. This section and the next section will introduce some of the additional insights which can be extracted once a classification model is constructed.

Institutional data is exceptionally complex; random forest classification
models allow the identification of the parts of the institutional data that
are important for the prediction of student risk and the thresholds in that data that go into
classifying a student as at-risk.

The first measure useful in further understanding which variables are most important
in the classification process is ``variable importance.'' The importance of a variable
to one of the model characterization metrics such as DFW accuracy is computed by
fitting the model with the variable and then without the variable to determine the
mean decrease in the characterization measure when the variable is removed from the model.
Figure \ref{fig:finalvar} shows the mean decrease in DFW accuracy, DFW precision, and overall
accuracy as the different variables used in the full model are removed for Sample 2 using data available in the second week of the class. Similar plots for Samples 1 and 3 are presented in the Supplemental Material \cite{supp}.

\begin{figure*}[!htb]
    \centering
    \includegraphics[width=5in]{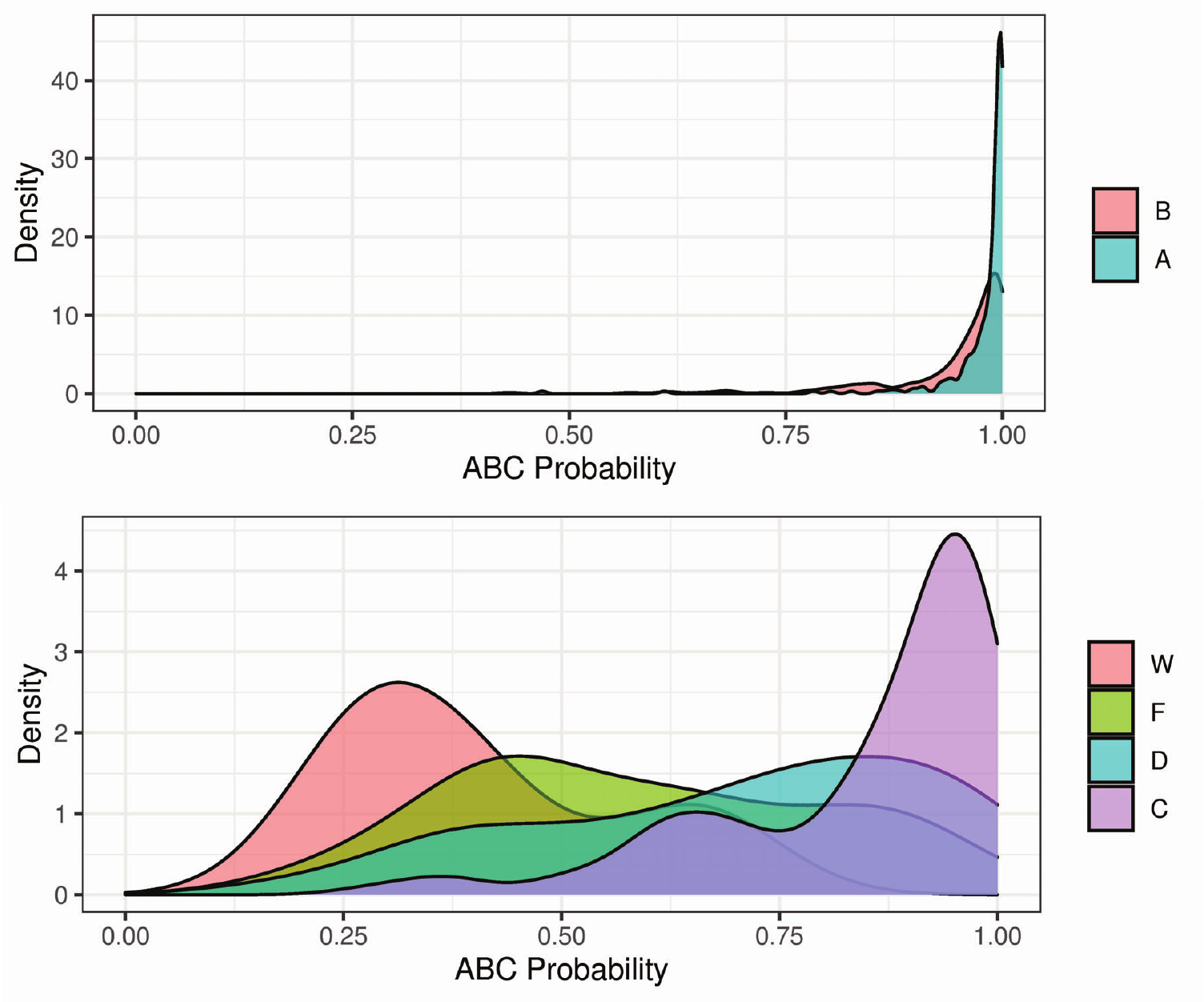}
    \caption{Predicted probability of earning an A, B, or C for Sample 1 disaggregated by the actual grade received in
    the class. The figure plots the probability density of each outcome. The order of the peaks in the lower figure from left to right is W, F, D, C. \label{fig:prob}   }
\end{figure*}

The variable importance plots shown in Fig. \ref{fig:finalvar}
show that homework average followed by CGPA were the most important variables in accurately identifying DFW students.
In addition to these variables, only CmpPct (the percentage of credit hours complete) has an error bar that does not include zero.  These results are very different than the variable importance results of Study 1 which predicted the AB outcome and used overall accuracy to measure model performance. In Study 1, while homework grade grew in variable importance from week to week, it was less important than CGPA until week 5 when test 1 was given. As
in Study 1, a very limited number of institutional variables were
needed to predict grades in a physics class.

While many instructors would select CGPA as an important variable
and would hope that homework averages were important,
quantitatively having a relative measure of importance is
valuable. The variable importance plots in Fig. \ref{fig:finalvar}
also identify many variables that seem important such as HSGPA,
ACTM, and demographic variables, which were not important for the
prediction of the DFW outcome.

\begin{figure*}[!htb]
    \centering
    \includegraphics[width=5in]{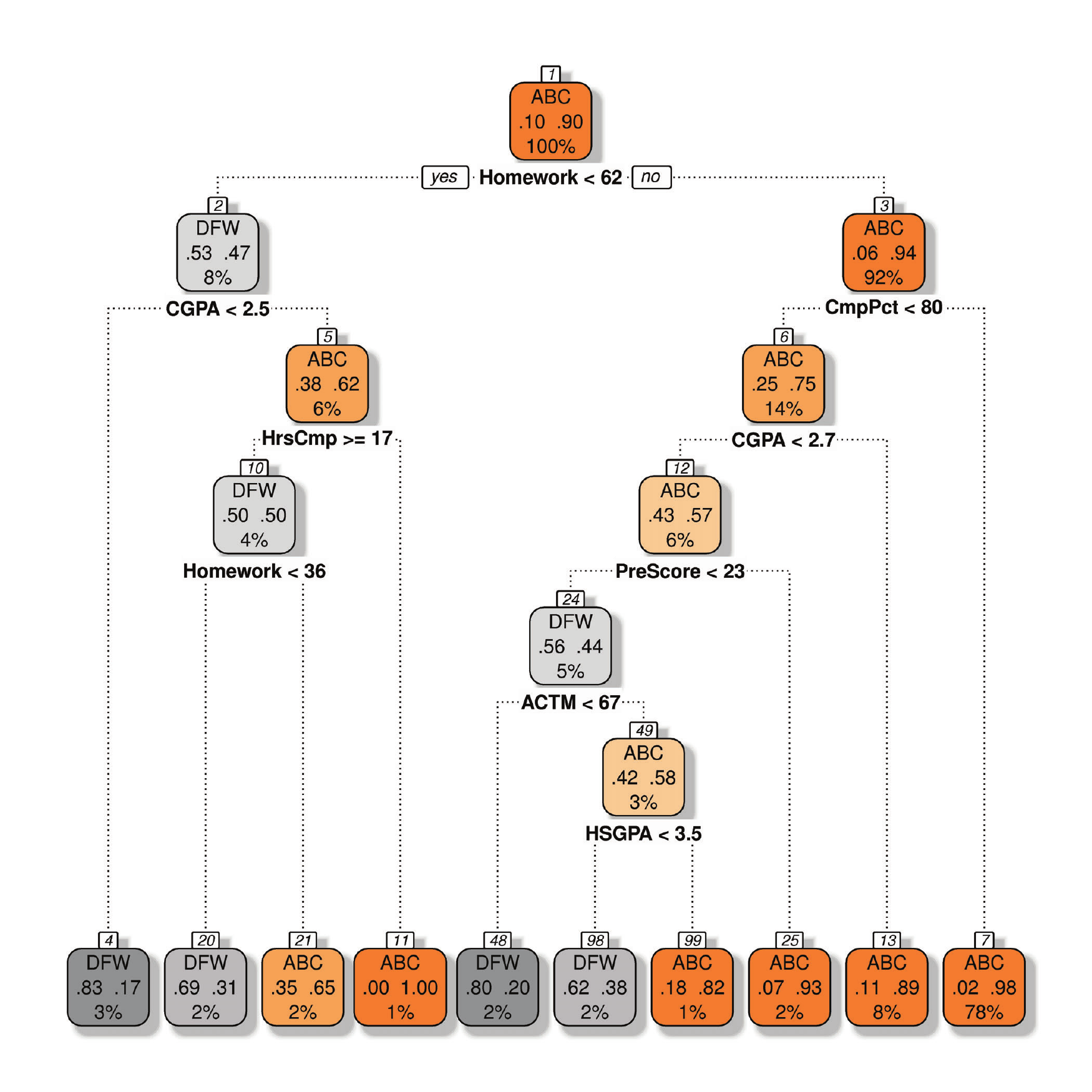}
    \caption{Decision tree for predicting the DFW outcome for Sample 2 using institutional data and
    data available in-class at the end of week 2. \label{fig:decisiontree}   }
\end{figure*}

\subsection{Applying Classification Models}
\label{sec:apply} The most basic output of a classification model
is the assignment of each student in the dataset into one of two
classes: those students likely to receive and A, B, or C and those
likely to receive a D, F, or W. Classification algorithms, once
constructed, can provide a finer grained picture of student risk
that may be more useful in applying machine learning results to
manage instructional interventions for at-risk students. A
classification model can also provide the probability a student
will receive each outcome. The predicted probability density
distribution of receiving an A, B, or C is plotted for each actual
grade outcome in Fig. \ref{fig:prob}. Two plots are provided to
improve readability. The distribution of probability estimates of
students who actually earn an A or B is very narrow with most
students with a predicted probability above 0.75. This suggests
that the students who actually receive A or B in the class are
predicted to receive an A, B, or C with very high probability. The
probability curve for students earning a C is much broader but
still peaked near one. Examination of the C distribution
illustrates two key features of the prediction: (1) the vast
majority of students who actually earn a C are predicted to do so
with probability $p>0.5$ and (2) some students who receive a C are
predicted to do so with very low probability. As such, an
instructor should not interpret a low probability of receiving a C
as a guarantee that a student will not succeed in the class. The
probability distributions of the F and W outcomes are very broad
showing these students are very difficult to predict accurately.
Examination of these distributions can help instructors understand
how an individual student's probability estimate translates into
actual grade outcomes and inform risk decisions.

Variable importance plots quantify the relative importance of the
many variables used in the classification model correcting for the
collinearity of many of the variables. These plots, however, do
not provide information about the levels of these variables
important in making the classification. A random forest grows
thousands of decision trees on a subset of the variables;
examining a single decision tree using all variables can show the
thresholds for the important variables. The decision tree for the
training dataset of Sample 2 in week 2 of the class is shown in
Fig. \ref{fig:decisiontree}. Each node in the tree is labeled with
the majority member of the node, either ABC or DFW. The root node
(top node) contains the entire training dataset,  indicated by the
100\%  at the bottom node. Every node indicates the fraction of
the training dataset contained in the node.  The fraction of each
outcome is shown in the center of the node; for example, the root
node contains 10\% DFW students and 90\% ABC students. The
decision condition is printed below the node. If the condition is
true for the student, the left branch of the tree is taken; if
false, the right branch is taken. For example, the decision
condition for the root node is whether the week 2 homework average
is above or below 62\%. For the 8\% of the students below this
average, the left branch is taken to node 2. Only 47\% of the
students in node 2 receive an A, C, or C. For the 3\% of these
students with CGPA less than 2.5, only 17\% receive an A, B, or C
(node 4). The decision tree gives a very clear picture of the
relative variable importance (higher variables in the tree are
more important) and the threshold of risk of receiving a D, F, or
W at each level of the tree.

\section{Discussion}
This study sought to answer three research questions; they will be addressed in the order proposed.

{\it RQ1: How can machine learning algorithms be applied to
predict unbalanced physics class outcomes?} Study 1 used random
forests and logistic regression to predict which students would
receive an A or B in introductory physics. The default random
forest parameters were used to build the models and the models
were characterized by their overall accuracy, $\kappa$, and AUC.
Because the outcome variable was fairly balanced, with 63\% of the
students receiving an A or B, overall accuracy provided an
acceptable measure of model performance. The pure guessing
accuracy was 63\%, and therefore, this statistic could vary over
the range 63\% to 100\% as variables were added to the model.

In the current work, the methods introduced in Study 1 were
unproductive because the outcome variable, predicting the DFW
outcome, was substantially unbalanced with only 10\% (Sample 2) to
20\% (Sample 3) of the students receiving this outcome. For this
outcome, the pure guessing overall accuracy (simply predicting
everyone receives an A, B, or C) is from 80\% to 90\% making it an
inappropriate statistic to judge model quality. This work
introduced the DFW accuracy and precision as more useful
statistics to evaluate model performance. In Sample 1, using the
default random forest algorithm parameters (Table
\ref{tab:results}, Default model) produced a model with very low
DFW accuracy identifying only 16\% of the students who actually
received a D, F, or W in the test dataset; however, 57\% of its
predictions were correct. This does not necessarily make it a bad
model, rather a model that is tuned for a specific purpose where
it is much more important for the predictions to be correct than
it is to identify the most potentially at-risk students. This
might be useful for an application that tries to identify students
for a high cost or non-negligible-risk intervention where only the
most likely at-risk students could be accommodated.

Multiple methods were explored to improve model performance:
oversampling, undersampling, hyperparameter tuning, and grid
search. This exploration is described in the Supplemental Material
\cite{supp}. All methods improved the balance of DFW accuracy and
precision. Oversampling led to models that overfit the data and
was not used. Grid search showed that, for this dataset, it was
always possible to use hyperparameter tuning by adjusting the
decision threshold without having to undersample to produce a
model with a balance of DFW accuracy and precision. The decision
threshold for models in Table \ref{tab:results} excluding the
default model and the models applied only to underrepresented
groups was adjusted for each model to balance DFW accuracy and
precision. For the overall model of Sample 1, this produced a
model with substantially higher DFW accuracy and $\kappa$ that the
default model; however, it still only identified 43\% of the
students who would receive a D, F, or W, DFW accuracy= 0.43, and
had $\kappa=0.36$ in the range fair agreement by Cohen's criteria.

Sample 2 restricted the time frame in which the institutional data
were collected to a 3-year period in which the course studied had
a consistent instructional environment. Even though the size of
the sample was much smaller, model performance was improved
showing that it is important to collect the training sample for a
period where the class was presented in the same form as the class
in which the model will be used.

The Sample 2 model using only institutional variables was much
better than models using only in-class variables early in the
semester. If an instructor wants to develop classification models
for prediction of students at risk early in the semester,
accessing a set of institutional data can substantially improve
the models. The combination of institutional and in-class
variables gave the highest model performance with an improvement
of 3\% in week 1, 6\% in week 2, 9\% in week 5 (when test 1 grades
were available), and 18\% in week 8 (when test 2 grades were
available) compared to the model containing only institutional
variables. As such, for identification of at-risk students early
in the semester most of the prediction accuracy can be achieved
with institutional data alone.

Sample 3 included a more restricted set of institutional variables
than Sample 1, but included a variable indicating socioeconomic
status and featured a more demographically diverse population. The
overall model for this sample had weaker performance metrics than
the overall model for Sample 1 or the institutional model for
Sample 2. When the set of variables used in Sample 1 was
restricted to be approximately those used in Sample 3, model
performance was commensurate. It is, therefore, important for
improving model performance to work with institutional research to
provide the machine learning algorithms with as rich a set of data
as possible.

{\it RQ2: Does classification accuracy differ for underrepresented
groups in physics? If so, how and why does it differ?} For Samples
1 and 3, once the model was constructed for the full training
dataset, the overall model was used to classify demographic
subgroups in the test dataset separately as shown in Table
\ref{tab:results}. These models examined women, URM students,
first-generation college students, and low SES students. In both
samples, the model performance metrics for some minority
demographic groups were different (either better or worse) than
the overall model; however, these differences were within one
standard deviation of the overall model. As such, the classifier
built on the full training dataset predicted the outcomes of
underrepresented physics students with approximately equal
accuracy. While the differences observed in Table
\ref{tab:results} are within the error of the sample, should
significant differences be detected, it is possible to re-tune the
models for each underrepresented group separately.

Figure \ref{fig:finalvar} and similar figures in the Supplemental
Material \cite{supp} show the demographic variables, gender, URM,
FirstGen, and SES are of low importance in the classification
models. This is likely because these factors already have a
general effect on other variables included in the models such as
CGPA. The Supplemental Material \cite{supp} includes an analysis
which undersamples the majority demographic class (for example,
men) to produce a more balanced dataset (for example, a dataset
with the same number of men and women) (Supplemental Figs. 7 to
9). The variable importance of the demographic variables used in
this study was fairly consistent with the rate of undersampling
showing that the low importance was not simply a result of the
lower number of students from minority demographic groups in the
sample.

To further investigate the low variable importance of the
demographic variables, we examined a more diverse population
(Sample 3). Model performance metrics were consistent with those
obtained from Sample 1, suggesting the low variable importance was
not the result of the restricted number of underrepresented
students in the sample.

{\it RQ3: How can the results of a machine learning analysis be
used to better understand and improve physics instruction?} Once a
classification model is constructed, the same model can be used to
characterize new groups of students. Sections \ref{sec:varimp} and
\ref{sec:apply} presented three different possible analyses that
can be performed with classification models that have classroom
applications.

The first analysis computed the variable importance of each
variable in the classifier, Fig. \ref{fig:finalvar}. This is done
by finding the mean decrease in some performance metric if the
variable is removed from the model. This analysis allows the
identification of the variables which are most predictive of a
student receiving a D, F, or W. This can show a working instructor
where to look in complex institutional datasets and allow
departments to shape their data requests.

The second analysis computed a probability of receiving an A, B,
or C for each individual student. This was plotted for each actual
grade received in Fig. \ref{fig:prob}. This allows an individual
quantitative risk to be applied to each student. This risk could
be updated as the semester progresses based on in-class
performance.

The final analysis computed a decision tree, Fig.
\ref{fig:decisiontree}. This tree shows the decision thresholds
which indicate the levels of the variable that are important in
classifying at-risk students. As long as the instructional setting
and assignment policy remains consistent, these trees can be
reused semester to semester without having to rerun the analysis.
The tree shows that homework average, CGPA, and the percent of
hours completed were important in the decision to classify a
student at risk of a DFW outcome.

These analysis results represent examples of the additional tools
classification algorithms can provide instructors; many more
examples could be given. The following represent some of the
applications of these results being considered at Institution 1.
These applications are designed around the principle that any
additional instructional activity must potentially benefit all
students. The models are far from perfect and, as such, all
students may actually be at risk so any intervention must be
available to any student.

{\bf Informing Resource Allocation:} Students in physics classes
at Institution 1 elect laboratory sections where a substantial
part of the interactive instruction in the course is presented.
Because a success probability can be generated for each student,
an average probability of success could be calculated for each
laboratory section. More experienced teaching assistants could be
assigned to these sections. The department also has a Learning
Assistant (LA) \cite{otero2010physics} program using a for-credit
model. Learning Assistants are not available for all lab sections;
allocation of LAs to at-risk lab sections could be prioritized.

{\bf Planning Revised Assignment Policy:} The decision tree in
Fig. \ref{fig:decisiontree} and variable importance measures in
Fig. \ref{fig:finalvar} shows that homework grades in the second
week of the class are the most important variable for predicting
success and gives a homework score threshold of 62\% as the
highest level decision for predicting success or failure. To
develop the habit of completing homework and investing sufficient
effort to do well on homework, a policy allowing the reworking of
homework assignments which received a grade of less than 60\% for
additional (or initial) credit could be implemented early in the
class.

{\bf Planning Student Communication:} Instructors can use the
variable importance results to provide general advice to students
with low homework grades and encourage them to seek additional
help by attending office hours or to change habits so homework
assignments are started earlier and sufficient time is allowed for
completion. In general, an instructor of a large service course
does not have time to personally communicate with each student;
however, the combination of the individual success probability,
variable importance, and variable decision threshold would allow
an instructor to monitor and communicate directly with a small
subset of students particularly at risk in the class. These
communications could let the students know that the instructor
noticed that early homework assignments needed additional work and
suggest strategies to the students for improvement opening
channels of personal communication with at-risk students.

Many other potential instructional uses of this type of analysis
are possible. Naturally, if the intervention is successful, it
will modify student outcomes changing students' risk profiles. The
classifier will need to be rebuilt using student outcomes after
the implementation of the intervention to reflect this modified
risk.

While using the random forest algorithm to make predictions is
technically fairly straightforward for instructors trained in
physics (the base code is presented in the Supplemental Material
\cite{supp}), obtaining the institutional dataset may present a
substantial barrier for overworked instructors of large service
introductory classes. As such, we present some recommendations for
managing the process of obtaining institutional data.

Gathering additional data for use by instructors should probably
be the responsibility of a departmental committee or staff. The
data required for different classes are quite similar. A
departmental data committee would also be able to establish
ethical standards for the use and handling of the data. Some
effort will be needed to understand the data available at the
institutional level and to work with institutional research to
fine tune the data request. For example, if one requests a basic
set of demographic and descriptive variables about students
enrolled in a course over a number of semesters, the GPA variable
provided will probably be the student's current GPA where one
actually wants the student's GPA before he or she enrolled in the
class of interest. Some interaction would also be required to
develop variables such as the student's math readiness or the
fraction of classes completed. However, once a set of variables is
identified, institutional records can quickly generate the data
for the department each semester. Once the institutional data is
acquired and understood, applying the machine learning code is
fairly straightforward. It is also worth pursuing the possibility
that institutional research could handle the entire process and
provide a machine learning risk analysis to interested
instructors. Student retention is of vital interest to most
institutions with retention in core mathematics and science
classes an important part of the puzzle.

\section{Ethical Considerations} The results of a machine
learning classification represent a new tool for physics
instructors to shape instruction; as with any tool, it can be
correctly used or misused. If an instructor is to use the
predictions of a classification algorithm, it is important that
these results do not bias his or her treatment of individual
students. Figure \ref{fig:prob} shows that it is possible for
students with very low predicted probability of earning an A, B,
or C to get a C or higher in the class. Machine learning
algorithms will never be 100\% accurate and this should be taken
into account in any application of the results of the algorithms.
Further, while the classification results may be used to direct
resources to the students most at risk, this should be done with
the goal of improving instruction for all students. Machine
learning results should also not be used to exclude students from
additional educational activities to support at-risk students.
Because the predictions are not 100\% accurate, additional
tutoring sessions or similar resources should be available to all;
however, the results of classification models could be used to
deliver encouragement to the students most at risk to avail
themselves of these opportunities.

\section{Conclusions}
This work applied the random forest machine learning algorithm to
predict whether introductory mechanics students would receive a
grade of D or F, or withdraw from a physics class. Metrics and
methods applied in previous work produced classification models
with poor performance; however, selecting metrics appropriate for
unbalanced outcomes and tuning the random forest models greatly
improved the classification accuracy of the DFW outcome.
Classification models performed similarly for students from two
institutions with very different demographic characteristics.
Models with a richer set of institutional variables were somewhat
(7\%) more accurate than models with a limited set of variables.
The addition of in-semester variables, particularly homework
averages and test scores, improved model performance. The
institutional model far outperformed a model using only
in-semester variables early in the semester; the performance of
the in-semester only models exceeded that of the institutional
only models once the first test was included as a variable. The
classifier trained on the full set of students produced somewhat
different performance for women, underrepresented minority
students, and first-generation college students with some metrics
improved and some weaker for these students. Once a classifier is
constructed, multiple new analyses are available allowing the
direction of additional resources to at-risk students.

\begin{acknowledgments}
    This work was supported in part by the National Science Foundation
    under grant ECR-1561517 and HRD-1834569.
\end{acknowledgments}


%

\end{document}